\documentclass[a4paper]{article}%
\usepackage{amsmath}%
\usepackage{amsfonts}%
\usepackage{amssymb}%
\usepackage{graphicx}
\begin{document}

\author{Sawa Manoff\\Bulgarian Academy of Sciences\\Institute for Nuclear Research and Nuclear Energy\\Department of Theoretical Physics\\Blvd. Tzarigradsko Chaussee 72\\1784 Sofia - Bulgaria}
\title{Variation of the velocity and the frequency of a periodic signal along the
world line of the emitter}
\date{E-mail address: smanov@inrne.bas.bg}
\maketitle

\begin{abstract}
The variation of the velocity of a periodic signal and its frequency along the
world line of a standard emitter (at rest with an observer) are considered in
a space with affine connections and metrics. It is shown that the frequency of
the emitted periodic signal is depending on the kinematic characteristics of
the motion of the emitter in space-time related to its shear and expansion
velocities. The same conclusions are valid for a standard clock moving with an observer.

PACS numbers: 95.30.Sf; 04.90.+h; 04.20.Cv; 04.90.+e

\textit{Short title}: Variation of velocity and frequency of a standard
periodic signal

\end{abstract}

\section{Introduction}

1. Modern problems of relativistic astrophysics as well as of relativistic
physics (dark matter, dark energy, evolution of the universe, measurement of
velocities of moving objects etc.) are related to the propagation of signals
in space or in space-time. The basis of experimental data received as results
of observations of the Doppler effect or of the Hubble effect gives rise to
considerations about the theoretical status of effects related to detection of
signals from emitters moving relatively to observers carrying detectors in
their laboratories. Nevertheless, in the last decades, there is no essential
evolution of the theoretical models related to new descriptions of the Doppler
and Hubble effects corresponding to the recent development of new mathematical
models for the space-time. In the astronomy and astrophysics standard
theoretical schemes for measuring velocities are used related to classical
mechanics and / or special and general relativity \cite{Lindegren},
\cite{Eva-Maria}.

2. The incoming periodic signals sent by an emitter moving relatively to an
observer (detector) are compared with periodic signals of an emitter lying at
rest with the observer. On this basis, the change of frequency and velocity of
the incoming periodic signals leads to conclusions about velocities and
accelerations of objects moving with respect to the observer.

\subsection{Standard periodic emitter}

1. A \textit{standard emitter} is an emitter moving together with an observer
(detector) in space-time and lying at rest in the proper frame of reference of
the observer (detector). The proper frame of reference of the standard emitter
could be identified with the proper frame of reference of the observer (detector).

2. In all considerations of effects related to the propagation of periodic
signals, it is assumed that an emitter is moving together with the observer,
i.e. an emitter is at rest in the proper frame of reference of the observer
(detector) and is used as a standard emitter with respect to which the
incoming signals from other emitters are compared with respect to their
velocity and frequency. On this basis, the variations of the frequency of
periodic signals of an emitter with relative velocity and accelerations with
respect to an observer (detector) could be established and the relative
velocity and acceleration could be measured. It is \textit{generally assumed}
that a standard emitter has constant frequency on its way in space-time
because of the fact that it is at rest with respect to the observer and no
changes based on Doppler effect could occur. This assumption is based on the
standard expression of the Doppler effect in classical mechanics and special
relativity, where no acceleration of the proper frame of the observer is taken
into account. But if the observer (detector) is moving with acceleration on
(non-geodesic in $V_{n}$-spaces or non-auto parallel) world line in space-time
the things could change and we could have results different from the standard
assumption for the constant frequency of an emitter moving in space-time
together with an observer. The question arises how the proper motion of a
standard emitter could influence the velocity and the frequency of its
periodic signals.

In this paper the variation of the absolute value of the velocity and the
frequency of signals sent by a standard emitter are considered in the proper
frame of reference of the emitter. In Section 2 the variation of the velocity
of a periodic signal along the world line of a standard emitter is considered.
In Section 3 the variation of the frequency of a periodic signal along the
world line of a standard emitter is determined. In Section 4 the variation of
the velocity and the frequency of periodic signals of a standard oscillator
(clock) are found. Some concluding remarks comprise the Section 5. It is shown
that the general belief that the frequency and the absolute value of the
velocity of periodic signals sent by a standard emitter does not change on the
world line of the emitter needs to be revised.

\subsection{Methods for determining space distances and relative velocities}

1. In the classical (non-quantum) field theories different models of
space-time have been used for description of the physical phenomena and their
evolution. The $3$ - dimensional Euclidean space $E_{3}$ is the physical space
used as the space basis of classical mechanics. The $4$ - dimensional (flat)
Minkowskian space $\overline{M}_{4}$ is used as the model of space-time in
special relativity. The (pseudo) Riemannian spaces $V_{4}$ without torsion are
considered as models of space-time in general relativity.. In theoretical
gravitational physics (pseudo) Riemannian spaces without torsion as well as
(pseudo) Riemannian spaces $U_{4}$ with torsion are proposed as space-time
grounds for new gravitational theories. To the most sophisticated models of
space-time belong the spaces with one affine connection and metrics
[$(L_{n},g)$-spaces] and the spaces with affine connections and metrics
[$(\overline{L}_{n},g)$-spaces] \cite{Manoff-1}, \cite{Manoff-2}.

2. All considerations related to the relative motions of objects with respect
to each other are made on the basis of the notions of relative velocity and
relative acceleration. The relative velocity is usually considered as radial
(centrifugal, centripetal) velocity and tangential (Coriolis) velocity.

There are two different approaches for measuring radial velocities of cosmic objects:

\begin{itemize}
\item a covariant approach related to proper times (\textit{spectroscopic
method}) and

\item a co-ordinate approach related to time and distances in a co-ordinate
system (identified as the proper frame of reference of the Sun)
(\textit{astrometric method}) \cite{Lindegren}.
\end{itemize}

The covariant method is usually specialized for a defined co-ordinate system.
This leads to difficulties related to the different definitions of
spectroscopic and kinematic quantities which are not directly connected to the
real measurement of the radial or tangential velocities of astronomical
objects. The comparison between the co-ordinate quantities and the
spectroscopic data leads to introduction of notions such as ''barycentric
radial-velocity measure'' and ''astrometric radial velocity'', where the
notion of real radial velocity is avoided. The reason for the introduction of
the above notions is the lack of covariant method for describing, on the one
side, the relative velocities and accelerations and, on the other side, the
lack of relations between these kinematic characteristics and their
corresponding Doppler shifts and Hubble shifts. Recently, it has been shown
that the introduced in $(\overline{L}_{n},g)$-spaces kinematic characteristics
related to relative velocities and relative accelerations could be in simple
way expressed in terms of radial (centrifugal, centripetal) and tangential
(Coriolis') velocities and accelerations \cite{Manoff-3}. These velocities and
accelerations could be considered as the velocities and accelerations of an
emitter from point of view of an observer.

\subsection{Periodic signals. Definitions and properties}

The notion of \textit{periodic signal} could be defined from physical and from
mathematical point of view.

\textit{From physical point of view} a periodic signal in a $(\overline{L}%
_{n},g)$-space, considered as a model of space-time, is characterized by:

\begin{itemize}
\item A periodic process, characterized by its \textit{direction
}and\textit{\ frequency}, transferred by an emitter and received by an
observer (detector).

\item A periodic process with finite velocity of propagation from point of
view of the observer, characterized by its \textit{absolute value of the
velocity} of propagation.
\end{itemize}

\textit{From mathematical point of view} a periodic signal in a $(\overline
{L}_{n},g)$-space, considered as a model of space-time, is characterized by:

\begin{itemize}
\item Isotropic (null) contravariant vector field $\widetilde{k}%
:g(\widetilde{k},\widetilde{k})=0$, $\widetilde{k}\in T(M)$, $dim\,M=n$,
$sgn\,g=n-2$ or $sgn\,g=-n+2$, determining the \textit{direction} of the
propagation of a periodic signal in space-time. $M$ is the differentiable
manifold with dimension $n$, provided with affine connections and metrics,
$T(M)$ is the tangent space over $M:T(M)=\cup_{x\in M}T_{x}(M)$.

\item Non-isotropic contravariant vector field $u:g(u,u)=e=\pm l_{u}^{2}\neq
0$.\thinspace The vector field $u\in T(M)$ is interpreted as the
\textit{velocity vector field of an observer (detector)}.

\item $l_{u}^{2}=\pm g(u,u)>0$, interpreted as the finite velocity of a
periodic signal, determining the \textit{absolute value of the velocity} of a
periodic signal with respect to the proper frame of reference of an observer.
The sign before $g(u,u)$ is depending on the signature of the metric of the space-time.

\item Scalar product of $\widetilde{k}$ and $u:g(\widetilde{k},u)=\omega>0$,
interpreted as the frequency of a periodic signal, determining the
\textit{frequency} of a periodic signal with respect to the proper frame of
reference of an observer (detector).
\end{itemize}

A \textit{frame of reference} is determined by the set of three geometric
objects \cite{Manoff-4}:

\begin{itemize}
\item A non-null (time like if $dim\,M=4$) contravariant vector field $u\in
T(M)$.

\item A tangent sub space $T_{x}^{\perp u}(M)$ orthogonal to $u$ at every
point $x\in M$, where $u$ is defined.

\item (Contravariant) affine connection $\nabla=\Gamma$. It determines the
type of transport along the trajectory to which $u$ is a tangent vector field.
$\Gamma$ is related to the covariant differential operator $\nabla_{u} $ along
$u$ \cite{Manoff-1}
\end{itemize}

Then the \textit{definition of a frame of reference} read

\textit{The set }$FR\sim[u$, $T^{\perp u}(M)$, $\nabla=\Gamma$, $\nabla_{u}]$
\textit{is called frame of reference in a differentiable manifold }%
$M$\textit{\ considered as a model of the space or of the space-time.}

\section{Variation of the velocity of a periodic signal along the world line
of a standard emitter}

Let us now consider the motion of a standard emitter moving with an observer.
First of all, let us recall the definitions for a periodic signal and its
theoretical description in $(\overline{L}_{n},g)$-spaces. Since all other
spaces, considered as mathematical models of a space or of a space-time, are
included in these types of spaces, all results would be valid for $(L_{n}%
,g)$-spaces, $U_{n}$-, $V_{n}$- spaces (spaces with affine connection and
metrics, (pseudo) Riemannian spaces with torsion, (pseudo) Riemannian spaces
without torsion) etc.

Let $\tau$ be the proper time (parameter) of the world line (trajectory)
$x^{i}(\tau)$, $i=1,...,n$, $dim\,M=n$, $n=4$ in a space-time of a standard
emitter [i.e. of an emitter lying at rest with an observer (detector))]. The
tangent vector field along the world line of the standard emitter could be
written in the form
\begin{equation}
u=\frac d{d\tau}=\frac{dx^{i}}{d\tau}\cdot\partial_{i}=u^{i}\cdot\partial
_{i}\text{ \thinspace\thinspace\thinspace\thinspace,\thinspace\thinspace
\thinspace\thinspace\thinspace\thinspace\thinspace\thinspace\thinspace
\thinspace\thinspace\thinspace}u^{i}=\frac{dx^{i}}{d\tau}%
\,\,\,\,\,\,\,\text{.}\label{1.1}%
\end{equation}

Its absolute length $l_{u}$ defined by the expression \cite{Manoff-1}
\begin{equation}
\pm l_{u}^{2}=g(u,u)=g_{\overline{i}\overline{j}}\cdot u^{i}\cdot
u^{j}\,\,\,\,\,\text{,\thinspace\thinspace\thinspace\thinspace\thinspace
\thinspace\thinspace\thinspace\thinspace\thinspace\thinspace\thinspace
\thinspace\thinspace\thinspace\thinspace\thinspace}g_{\overline{i}\overline
{j}}=f^{k}\,_{i}\cdot f^{l}\,_{j}\cdot g_{kl}\text{ \thinspace\thinspace
\thinspace\thinspace,}\label{1.2}%
\end{equation}
where
\begin{align}
g  & =g_{ij}\cdot dx^{i}.dx^{j}\text{ \thinspace\thinspace\thinspace
\thinspace\thinspace\thinspace,\thinspace\thinspace\thinspace\thinspace
\thinspace\thinspace\thinspace\thinspace\thinspace\thinspace\thinspace
\thinspace}dx^{i}.dx^{j}=\frac12\cdot(dx^{i}\otimes dx^{j}+dx^{j}\otimes
dx^{i})\,\,\,\text{,}\label{1.3}\\
f^{k}\,_{i}  & =f^{k}\,_{i}(x^{l})\in C^{\infty}(M)\text{ .}\nonumber
\end{align}
could be interpreted as the absolute value of the velocity of the periodic
signal, measured in the proper frame of reference of the observer (detector)
now identified with the proper frame of reference of the standard emitter.

The \textit{space direction} of the propagation of a periodic signal is given
by the contravariant vector field $k_{\perp}$%
\begin{equation}
k_{\perp}=\overline{g}[h_{u}(\widetilde{k})]=g^{ij}\cdot h_{\overline
{j}\overline{k}}\cdot\widetilde{k}\,^{k}\cdot\partial_{i}\text{ \thinspace
\thinspace\thinspace\thinspace\thinspace\thinspace\thinspace,}\label{1.4}%
\end{equation}
where
\begin{align*}
g(\widetilde{k},\widetilde{k})  & =0\text{ \thinspace\thinspace\thinspace
\thinspace\thinspace\thinspace\thinspace,\thinspace\thinspace\thinspace
\thinspace\thinspace\thinspace\thinspace\thinspace\thinspace\thinspace
\thinspace\thinspace\thinspace\thinspace\thinspace\thinspace}\overline
{g}=g^{ij}\cdot\partial_{i}.\partial_{j}\,\,\,\,\,\,\,\text{,}\\
\partial_{i}.\partial_{j}\,\,  & =\frac12\cdot(\partial_{i}\otimes\partial
_{j}+\partial_{j}\otimes\partial_{i})\,\,\text{\thinspace\thinspace
\thinspace\thinspace\thinspace\thinspace\thinspace\thinspace,}\\
h_{u}  & =g-\frac1{g(u,u)}\cdot g(u)\otimes g(u)\text{ \thinspace
\thinspace\thinspace\thinspace,}\\
g(u,u)  & =\pm l_{u}^{2}=e\neq0\,\,\,\,\,\,\,\text{.}%
\end{align*}

The contravariant isotropic (null) vector field $\widetilde{k}$ determines the
\textit{direction} of the periodic signal \textit{in the space-time}.

The frequency $\omega$ of the periodic standard signal (the periodic signal
sent by a standard emitter) is determined by the relations \cite{Manoff-5}:
\begin{equation}
\omega=g(u,\widetilde{k})=l_{u}\cdot g(\widetilde{n}_{\perp},k_{\perp
})\,\,\,\,\,\,\text{,}\label{1.5}%
\end{equation}
where
\begin{equation}
k_{\perp}=\mp l_{k_{\perp}}\cdot\widetilde{n}_{\perp}=\mp\frac\omega{l_{u}%
}\cdot\widetilde{n}_{\perp}\text{ \thinspace\thinspace\thinspace
\thinspace,\thinspace\thinspace\thinspace\thinspace\thinspace\thinspace
\thinspace\thinspace\thinspace\thinspace\thinspace}g(\widetilde{n}_{\perp
},\widetilde{n}_{\perp})=\mp1\text{ \thinspace\thinspace\thinspace
.}\label{1.6}%
\end{equation}

Once more, a standard emitter is an emitter staying at rest with respect to an
observer (detector). This means that the world lines of emitter and observer
are identical in the frame of reference of the observer (emitter). In other
words, the emitter and the frequency meter are at rest to each other and are
moving together in space-time.

The notion of a clock is closely related to the notion of a periodic signal.
''A clock is a physical device consisting of an oscillator running at some
angular frequency $\omega$ and a counter that counts the cycles. The period of
the oscillator, $T=2\pi/\omega$, is calibrated in some standard oscillator.
The counter simply counts the cycles of the oscillator. Since some epoch, or
the event at which the count started, we say that a quantity of time equal to
$NT$ has elapsed, if $N$ cycles have been counted'' \cite{Bahder}.

Let us now consider the variation of the absolute value $l_{u}$ of the
velocity of a periodic signal of a standard emitter. For this purpose we
consider the change of $l_{u}^{2}$ along the vector field $u$. Since
$l_{u}^{2}=\pm g(u,u)$, we have the relations
\begin{align}
\nabla_{u}(l_{u}^{2})  & =u(l_{u}^{2})=2\cdot l_{u}\cdot u(l_{u})=2\cdot
l_{u}\cdot\frac{dl_{u}}{d\tau}=\nonumber\\
& =\pm\nabla_{u}[g(u,u)]=\pm[(\nabla_{u}g)(u,u)+2\cdot g(u,a)]\text{
\thinspace\thinspace\thinspace,}\label{1.7}\\
a  & =\nabla_{u}u\text{ \thinspace\thinspace\thinspace\thinspace.}\nonumber
\end{align}
where $a=\nabla_{u}u$ is the acceleration of the standard emitter along its
world line, i.e. $a$ is the deviation of the world line from the corresponding
auto-parallel world line $\nabla_{u}u=0$. Therefore,
\begin{equation}
l_{u}\cdot\frac{dl_{u}}{d\tau}=\pm[g(u,a)+\frac12\cdot(\nabla_{u}%
g)(u,u)]\,\,\,\,\,\,\,\text{.}\label{1.8}%
\end{equation}

\textit{Special case:} $\overline{U}_{n}$-, $U_{n}$-, $\overline{V}_{n}$-, and
$V_{n} $-spaces. $\nabla_{u}g=0$ for $\forall u\in T(M)$.
\begin{align*}
l_{u}\cdot\frac{dl_{u}}{d\tau}  & =\pm g(u,a)=\pm g(u,a_{\parallel
})\,\,\,\,\,\,\text{,}\\
a_{\parallel}  & =\frac1e\cdot g(u,a)\cdot u\,\,\,\,\,\,\,\,\text{,\thinspace
\thinspace\thinspace\thinspace\thinspace\thinspace\thinspace\thinspace
\thinspace\thinspace}e=g(u,u)\text{ \thinspace\thinspace\thinspace
\thinspace\thinspace.}%
\end{align*}

\textit{Special case:} General relativity in $V_{n}$-spaces. In general
relativity the absolute value of a light signal is normalized to $l_{u}=c=$
const., or $l_{u}=1$. Then $g(u,a)=g(u,a_{\parallel})=0$. The acceleration
$a=a_{\perp}=\overline{g}[h_{u}(a)]$ is orthogonal to the vector field $u$. It
is lying in the sub space orthogonal to $u$. There are two reasons for the
normalization of $u$: one is from mathematical point of view, and the other is
from physical point of view. From mathematical point of view, every non-null
(non-isotropic) contravariant vector field $u$ could be normalized by the use
of the absolute value $l_{u}$ of its length in the form
\begin{align*}
\overline{u}  & =\pm\cdot\frac{c_{0}}{l_{u}}\cdot u\text{ \thinspace
\thinspace\thinspace, \thinspace\thinspace\thinspace\thinspace\thinspace
\thinspace\thinspace\thinspace\thinspace}g(\overline{u},\overline{u}%
)=\frac{c_{0}^{2}}{l_{u}^{2}}\cdot g(u,u)=\pm c_{0}^{2}\text{ \thinspace
\thinspace,}\\
c_{0}  & =\text{const. }\neq0\text{ .}%
\end{align*}

From physical point of view, \textit{it is assumed} that a light signal is
propagating with constant absolute value $l_{u}=$ const. from point of view of
the frame of reference of an observer. This point of view leads to its
mathematical realization by means of the normalization of the vector field $u
$.

\textit{Special case:} Auto-parallel motion of a standard emitter. This type
of motion is described by the equations
\begin{align*}
a  & =f\cdot u\,\,\,\,\,\,\,\text{,\thinspace\thinspace\thinspace
\thinspace\thinspace\thinspace\thinspace\thinspace\thinspace\thinspace
\thinspace}f\in C^{r}(M)\,\,\,\,\,\,\text{,\thinspace\thinspace\thinspace
\thinspace\thinspace\thinspace\thinspace\thinspace\thinspace\thinspace
\thinspace}r\geqslant2\text{ \thinspace\thinspace\thinspace\thinspace,}\\
a  & =0\,\,\,\,\,\,\text{.}%
\end{align*}

In the case $a=f\cdot u$, it follows that
\[
\frac1{l_{u}}\cdot\frac{dl_{u}}{d\tau}=f\pm\frac1{2\cdot l_{u}^{2}}%
\cdot(\nabla_{u}g)(u,u)\,\,\,\,\,\text{.}
\]

In the case $a=0$, it follows that
\[
\frac1{l_{u}}\cdot\frac{dl_{u}}{d\tau}=\pm\frac1{2\cdot l_{u}^{2}}\cdot
(\nabla_{u}g)(u,u)\,\,\,\,\,\text{.}
\]

\textit{Special case:} Geodesic motion of a standard emitter in general
relativity in $V_{n}$-spaces. Since $\nabla_{u}g=0$, the auto-parallel
(geodesic) motion described by $a=f\cdot u$ or $a=0$ leads to the relations
\begin{align*}
\frac1{l_{u}}\cdot\frac{dl_{u}}{d\tau}  & =f\,\,\,\,\,\,\,\,\,\,\,\text{,}\\
\frac1{l_{u}}\cdot\frac{dl_{u}}{d\tau}  & =0\,\,\,\,\,\,\,\text{.}%
\end{align*}

If the term $f\cdot u$ is interpreted as a type of friction then in a $V_{n}%
$-space we have the relation
\[
l_{u}=l_{u0}\cdot exp(f\cdot d\tau)\text{ \thinspace\thinspace\thinspace
\thinspace,\thinspace\thinspace\thinspace\thinspace\thinspace\thinspace
\thinspace\thinspace\thinspace\thinspace\thinspace\thinspace\thinspace
\thinspace\thinspace\thinspace}l_{u0}=\,\text{const.}
\]

If $f>0$, $l_{u}$ will increase with the time. If $f<0$, $l_{u}$ will decrease
with the time. Therefore, if we use a geodesic equation in its non-canonical
form $(a=0)$ then the absolute value of the velocity of a standard periodic
signal would change during a time period, where the proper time of the world
line of the standard emitter does not appear as the affine parameter of the
geodesic trajectory.

In the case $a=0$, it follows that
\begin{align*}
\frac1{l_{u}}\cdot\frac{dl_{u}}{d\tau}  & =0\text{ \thinspace\thinspace
\thinspace\thinspace\thinspace\thinspace\thinspace\thinspace\thinspace
\thinspace\thinspace, \thinspace\thinspace\thinspace\thinspace\thinspace
\thinspace\thinspace\thinspace}l_{u}\neq0\,\,\,\,\,\,\,\text{,}\\
\frac{dl_{u}}{d\tau}  & =0\,\,\,\,\,\,\,\,\,\,\,\,\text{,\thinspace
\thinspace\thinspace\thinspace\thinspace\thinspace\thinspace\thinspace
\thinspace\thinspace\thinspace}l_{u}=l_{u0}=\,\text{const. }\neq0\text{ .}%
\end{align*}

Therefore, the absolute value $l_{u}$ of a standard periodic signal is a
constant quantity along the geodesic world line of the standard emitter. This
means that if a standard emitter is moving on a geodesic trajectory in a
(pseudo) Riemannian space without torsion then the absolute value\thinspace
$l_{u}$ of its periodic signals will be a constant quantity along this
geodesic world line, i.e. \thinspace$l_{u}=l_{u0}=\,$const. $\neq0 $.

\textit{Special case:} Weyl's spaces with torsion ($Y_{n}$-spaces, Weyl-Cartan
spaces). For this type of spaces the condition
\[
\nabla_{u}g=\frac1n\cdot Q_{u}\cdot g\text{ \thinspace\thinspace
\thinspace\thinspace\thinspace\thinspace\thinspace\thinspace,\thinspace
\thinspace\thinspace\thinspace\thinspace\thinspace\thinspace\thinspace
\thinspace\thinspace\thinspace\thinspace\thinspace\thinspace\thinspace
\thinspace}dim\,M=n\text{\thinspace\thinspace\thinspace\thinspace,}
\]
is fulfilled. Then
\[
\frac1{l_{u}}\cdot\frac{dl_{u}}{d\tau}=\frac1{2\cdot n}\cdot Q_{u}\pm
\frac1{l_{u}^{2}}\cdot g(u,a)\,\,\,\,\,\,\text{.}
\]

\textit{Special case:} Auto-parallel motion of a standard emitter in Weyl's
spaces with torsion.
\[
\nabla_{u}g=\frac1n\cdot Q_{u}\cdot g\text{ \thinspace\thinspace
\thinspace\thinspace\thinspace\thinspace,\thinspace\thinspace\thinspace
\thinspace\thinspace\thinspace\thinspace\thinspace\thinspace\thinspace
\thinspace\thinspace\thinspace}a=0\,\,\,\,\,\,\,\,\text{,}
\]
\[
l_{u}=(l_{u0}^{2}+\frac1n\cdot\int Q_{u}\cdot d\tau)^{1/2}\text{
\thinspace\thinspace\thinspace\thinspace\thinspace\thinspace\thinspace
\thinspace,\thinspace\thinspace\thinspace\thinspace\thinspace\thinspace
\thinspace\thinspace\thinspace}l_{u0}^{2}=\text{ const. }>0\text{
\thinspace\thinspace.}
\]

Therefore, if a standard emitter is moving in a Weyl's space with torsion on
an auto-parallel world line then the absolute value of the velocity of its
periodic signals would change with the time under the existence of the
quantity $Q_{u}$ as a characteristic of this type of spaces. If, further,
$Q_{u}=-n\cdot d\overline{\varphi}/d\tau$, [$\overline{\varphi}=\overline
{\varphi}(x^{i}(\tau))=\overline{\varphi}(\tau)$ is a dilaton field in a
$Y_{n}$-space], then
\[
l_{u}=[l_{u}^{2}-\overline{\varphi}(\tau)]^{1/2}%
\,\,\,\,\,\,\,\,\text{,\thinspace\thinspace\thinspace\thinspace\thinspace
\thinspace\thinspace\thinspace\thinspace\thinspace\thinspace\thinspace}%
l_{u0}^{2}=\text{ const. }>0\text{ \thinspace\thinspace.\thinspace
\thinspace\thinspace\thinspace}
\]

The dilaton field $\overline{\varphi}$ influences the absolute value of the
velocity of a periodic signal emitted by a standard emitter. This fact could
be used as a check-up of the existence of a dilaton field in a Weyl-Cartan
space considered as a model of space-time.

In the further considerations the question could arise under which conditions
a standard emitter could send periodic signals with constant absolute value of
its velocity along its world line, measured in the proper frame of reference
of the emitter.

If we consider the conditions under which the absolute value $l_{u}$ of the
velocity of a periodic signal appears as a conserved quantity in a
$(\overline{L}_{n},g)$-space, we can prove the following propositions:

\textit{Proposition 1.} The necessary and sufficient condition for the
absolute value $l_{u}$ of the velocity of a periodic signal to be a conserved
quantity along the world line of a standard emitter with tangent vector field
$u$ is the condition
\begin{equation}
g(u,a)=-\frac12\cdot(\nabla_{u}g)(u,u)\,\,\,\,\,\,\,\,\,\text{,}\label{1.9}%
\end{equation}
equivalent to the condition
\begin{equation}
(\pounds _{u}g)(u,u)=0\text{ .}\label{1.10}%
\end{equation}

The proof is trivial. One should only take into account the relations
\begin{align*}
l_{u}\cdot\frac{dl_{u}}{d\tau}  & =\pm[g(u,a)+\frac12\cdot(\nabla
_{u}g)(u,u)]\,\,\text{\thinspace\thinspace\thinspace\thinspace\thinspace
,\thinspace\thinspace\thinspace\thinspace\thinspace\thinspace\thinspace
\thinspace\thinspace\thinspace\thinspace}l_{u}\neq0\text{ \thinspace
\thinspace,}\\
\pounds _{u}[g(u,u)]  & =u[g(u,u)]=\nabla_{u}[g(u,u)]=(\pounds _{u}%
g)(u,u)\text{ \thinspace\thinspace.}%
\end{align*}

\textit{Special case:} $V_{n}$-spaces as models of space-time in general
relativity theory. Since $\nabla_{u}g=0$, we have as necessary and sufficient
conditions the conditions
\[
g(u,a)=0\,\,\,\,\,\,\,\triangleq\,\,\,\,\,\,\,\,\,\,\,\,\,\,(\pounds
_{u}g)(u,u)=0\text{ .}
\]

\textit{Proposition 2.} A sufficient condition for the absolute value
$l_{u}\neq0$ of the velocity of a periodic signal of an emitter to be a
constant quantity $(l_{u}=$ const. $\neq0)$ along the world line of the
emitter with tangent vector field $u$ is the condition
\begin{equation}
\pounds _{u}g=0\,\,\,\,\,\,\,\text{,}\label{1.11}%
\end{equation}
i.e. if the vector field $u$ is tangent vector field to the world line of a
standard emitter then the absolute value $l_{u}$ of the velocity of its
periodic signals is a constant quantity $(l_{u}=$ const. $\neq0)$ along the
world line of the emitter when $u$ is a Killing vector field.

The proof is trivial.

\textit{Special case}: $V_{n}$-spaces as models of space-time in general
relativity theory.

(a) A sufficient condition for $l_{u}=$ const. $\neq0$ is the condition for
the world line of the standard emitter to be a geodesic trajectory in a given
$V_{n}$-space, i.e. $\nabla_{u}u=a=0$.

(b) A sufficient condition for $l_{u}=$ const. $\neq0$ is the vector field $u$
as tangent vector field to the world line of the standard emitter to be a
Killing vector field, i.e. $\pounds _{u}g=0\,$.

\section{Variation of the frequency of a periodic signal along the world line
of an emitter}

If an emitter is used as a standard emitter by an observer on his world line
for comparison of the incoming periodic signals from another emitters then the
variation of the frequency of the standard emitter is of great importance for
the correct determination of the frequency of the incoming signals.

From the explicit form of the frequency $\omega$%
\begin{equation}
\omega=l_{u}\cdot g(\widetilde{n}_{\perp},k_{\perp})\,\,\,\,\,\text{,}%
\label{2.1}%
\end{equation}
where \cite{Manoff-5}
\begin{equation}
k_{\perp}=\mp\frac\omega{l_{u}}\cdot\widetilde{n}_{\perp}%
\,\,\,\,\,\,\,\text{,\thinspace\thinspace\thinspace\thinspace\thinspace
\thinspace\thinspace\thinspace\thinspace\thinspace\thinspace\thinspace
\thinspace}g(\widetilde{n}_{\perp},u)=0\text{\thinspace\thinspace
\thinspace\thinspace\thinspace\thinspace,}\label{2.2}%
\end{equation}
it follows that
\begin{align}
\nabla_{u}\omega & =u\omega=\frac{d\omega}{d\tau}=\nabla_{u}[l_{u}\cdot
g(\widetilde{n}_{\perp},k_{\perp})\,]=\nonumber\\
& =ul_{u}\cdot g(\widetilde{n}_{\perp},k_{\perp})+\nonumber\\
& \,+l_{u}\cdot[(\nabla_{u}g)(\widetilde{n}_{\perp},k_{\perp})\,+g(\nabla
_{u}\widetilde{n}_{\perp},k_{\perp})+g(\widetilde{n}_{\perp},\nabla
_{u}k_{\perp})]\text{ \thinspace\thinspace.}\label{2.3}%
\end{align}

\textit{Remark.} If we use the relations
\begin{align}
l_{u}  & =\lambda\cdot\nu=\lambda\cdot\frac\omega{2\cdot\pi}%
\,\,\,\,\,\,\,\,\text{,}\label{2.4}\\
\omega & =2\cdot\pi\cdot\frac{l_{u}}\lambda=l_{u}\cdot g(\widetilde{n}_{\perp
},k_{\perp})\,\,\,\,\,\,\,\text{,}\label{2.5}%
\end{align}
we can find the expression for the corresponding to $\omega$ length $\lambda$
of the periodic signal in the form
\begin{equation}
\lambda=\frac{2\cdot\pi}{g(\widetilde{n}_{\perp},k_{\perp})}%
\,\,\,\,\,\,\,\,\,\,\text{,}\label{2.6}%
\end{equation}
and therefore,
\begin{equation}
g(\widetilde{n}_{\perp},k_{\perp})=\frac{2\cdot\pi}\lambda\,\,\,\,\,\,\text{.}%
\label{2.7}%
\end{equation}

The projection of $k_{\perp}$ on its unit vector $\widetilde{n}_{\perp}$ has
exact relation to the length $\lambda$ of the periodic signal with frequency
$\omega$.

After straightforward computations we obtain the following relation
\begin{equation}
h_{u}(\pounds _{u}\widetilde{n}_{\perp},\widetilde{n}_{\perp})+d(\widetilde
{n}_{\perp},\widetilde{n}_{\perp})+\frac12\cdot(\nabla_{u}g)(\widetilde
{n}_{\perp},\widetilde{n}_{\perp})=0\,\,\,\,\text{,}\label{2.8}%
\end{equation}
where $d$ is the deformation velocity tensor
\begin{equation}
d=\sigma+\omega+\frac1{n-1}\cdot\theta\cdot h_{u}\text{ \thinspace
\thinspace\thinspace.}\label{2.9}%
\end{equation}

The trace free covariant symmetric tensor $\sigma$ is the shear velocity
tensor. The antisymmetric covariant tensor $\omega$ is the rotation velocity
tensor. The invariant $\theta$ is the expansion velocity invariant
\cite{Stephani}, \cite{Manoff-1}.

The vector $k_{\perp}$ is directed in the line of sight, and the vector field
$u$ is orthogonal to it. The vector field $u$ is a tangent vector field to the
world line of the standard emitter. If we wish to consider the world line of
the emitter and the line of sight as co-ordinates lines along which we
consider the propagation of the periodic signals (the absolute value of their
velocity and direction) then we should admit the validity of the relations
(the second condition is fulfilled by the construction of $k_{\perp}$)
\begin{equation}
\pounds _{u}k_{\perp}=0\,\,\,\,\,\,\,\,\,\text{,\thinspace\thinspace
\thinspace\thinspace\thinspace\thinspace\thinspace\thinspace\thinspace
\thinspace\thinspace\thinspace\thinspace\thinspace\thinspace\thinspace
\thinspace\thinspace}g(u,k_{\perp})=0\,\,\,\,\,\,\,\text{.}\label{2.10}%
\end{equation}

\textit{Remark.} The above relations are the necessary and sufficient
conditions for the existence of co-ordinates curves to which $u$ and
$k_{\perp}$ appear as tangent vectors at every point of the corresponding
curve \cite{Bishop}.

By the use of the expressions
\begin{align*}
\pounds _{u}k_{\perp}  & =\pounds _{u}(\mp\frac\omega{l_{u}}\cdot\widetilde
{n}_{\perp})=\\
& =\mp[\frac d{d\tau}(\frac\omega{l_{u}})]\cdot\widetilde{n}_{\perp}\mp
\frac\omega{l_{u}}\cdot\pounds _{u}\widetilde{n}_{\perp}\,=0\,\,\,\,\text{,}%
\end{align*}
\[
\pounds _{u}\widetilde{n}_{\perp}=-[\frac d{d\tau}(log\frac\omega{l_{u}%
})]\cdot\widetilde{n}_{\perp}\,\,\,\,\,\,\,\,\,\,\,\,\text{,\thinspace
\thinspace\thinspace\thinspace\thinspace\thinspace\thinspace\thinspace
\thinspace\thinspace\thinspace\thinspace\thinspace\thinspace}u=\frac d{d\tau
}\,\,\,\,\,\,\,\text{,}
\]
\[
h_{u}(\pounds _{u}\widetilde{n}_{\perp},\widetilde{n}_{\perp})+d(\widetilde
{n}_{\perp},\widetilde{n}_{\perp})+\frac12\cdot(\nabla_{u}g)(\widetilde
{n}_{\perp},\widetilde{n}_{\perp})=0\,\,\,\,\text{,}
\]
after substitution of the explicit form for $\pounds _{u}\widetilde{n}_{\perp}
$ in the last above expression, the relation follows
\begin{equation}
-[\frac d{d\tau}(log\frac\omega{l_{u}})]\cdot h_{u}(\widetilde{n}_{\perp
},\widetilde{n}_{\perp})+d(\widetilde{n}_{\perp},\widetilde{n}_{\perp}%
)+\frac12\cdot(\nabla_{u}g)(\widetilde{n}_{\perp},\widetilde{n}_{\perp
})=0\text{ \thinspace\thinspace.}\label{2.11}%
\end{equation}

\subsection{Explicit form of the equation for $\omega$}

Since
\begin{equation}
h_{u}(\widetilde{n}_{\perp},\widetilde{n}_{\perp})=g(\widetilde{n}_{\perp
},\widetilde{n}_{\perp})=\mp1\text{ \thinspace\thinspace\thinspace
\thinspace\thinspace,}\label{2.12}%
\end{equation}
we have
\[
\frac d{d\tau}(log\frac\omega{l_{u}})=\mp[d(\widetilde{n}_{\perp}%
,\widetilde{n}_{\perp})+\frac12\cdot(\nabla_{u}g)(\widetilde{n}_{\perp
},\widetilde{n}_{\perp})]\,\,\,\,\,\,\,\text{.}
\]

On the other side,
\begin{align}
d(\widetilde{n}_{\perp},\widetilde{n}_{\perp})  & =(\sigma+\omega+\frac
1{n-1}\cdot\theta\cdot h_{u})(\widetilde{n}_{\perp},\widetilde{n}_{\perp
})=\nonumber\\
& =\mp[\frac1{n-1}\cdot\theta\mp\sigma(\widetilde{n}_{\perp},\widetilde
{n}_{\perp})]\,\,\,\,\,\,\text{,}\label{2.13}%
\end{align}
\begin{equation}
\frac1{n-1}\cdot\theta\mp\sigma(\widetilde{n}_{\perp},\widetilde{n}_{\perp
})=H\,\,\,\,\,\,\,\,\,\,\,\,\text{,\thinspace\thinspace\thinspace
\thinspace\thinspace\thinspace\thinspace\thinspace\thinspace\thinspace
\thinspace\thinspace\thinspace\thinspace}d(\widetilde{n}_{\perp},\widetilde
{n}_{\perp})=\mp H\,\,\,\,\,\,\text{.\thinspace\thinspace}\label{2.14}%
\end{equation}

The function $H=H(\tau)$ is the s.c. Hubble function \cite{Manoff-3},
\cite{Manoff-5}. Therefore, we can write now the expression
\begin{equation}
\frac d{d\tau}(log\frac\omega{l_{u}})=H\mp\frac12\cdot(\nabla_{u}%
g)(\widetilde{n}_{\perp},\widetilde{n}_{\perp})\text{ \thinspace
\thinspace\thinspace\thinspace\thinspace.}\label{2.15}%
\end{equation}

If we introduce the abbreviation
\begin{align}
\overline{V}(\widetilde{n}_{\perp},\widetilde{n}_{\perp})  & =\mp
[d+\frac12\cdot(\nabla_{u}g)](\widetilde{n}_{\perp},\widetilde{n}_{\perp
})=\label{2.16}\\
& =H\mp\frac12\cdot(\nabla_{u}g)(\widetilde{n}_{\perp},\widetilde{n}_{\perp
})\text{ ,}\label{2.17}\\
\overline{V}  & =\mp[d+\frac12\cdot(\nabla_{u}g)]\text{ \thinspace
\thinspace\thinspace\thinspace\thinspace,}\label{2.18}%
\end{align}
we obtain the equation for $\omega$%
\begin{equation}
\frac d{d\tau}(log\frac\omega{l_{u}})=\overline{V}(\widetilde{n}_{\perp
},\widetilde{n}_{\perp})\,\,\,\,\,\text{.}\label{2.19}%
\end{equation}

Its solution follows in the form
\begin{align*}
\frac\omega{l_{u}}  & =\frac{\omega_{0}}{l_{u0}}\cdot exp(\int[H\mp
\frac12\cdot(\nabla_{u}g)(\widetilde{n}_{\perp},\widetilde{n}_{\perp})]\cdot
d\tau)\text{ \thinspace\thinspace\thinspace\thinspace\thinspace,}\\
\omega_{0}  & =\text{ const.,\thinspace\thinspace\thinspace\thinspace
\thinspace\thinspace\thinspace\thinspace\thinspace\thinspace\thinspace
\thinspace\thinspace\thinspace\thinspace}l_{u0}=\text{ const.}%
\end{align*}

Since
\begin{equation}
\omega=2\cdot\pi\cdot\frac{l_{u}}\lambda\,\,\,\,\,\,\,\,\text{,\thinspace
\thinspace\thinspace\thinspace\thinspace\thinspace\thinspace\thinspace
\thinspace\thinspace\thinspace\thinspace}\frac\omega{l_{u}}=\frac{2\cdot\pi
}\lambda\,\,\,\,\,\,\,\,\text{,\thinspace\thinspace\thinspace\thinspace
\thinspace\thinspace\thinspace\thinspace\thinspace\thinspace\thinspace
\thinspace\thinspace\thinspace}l_{u}=\frac{\omega\cdot\lambda}{2\cdot\pi
}\,\,\,\,\,\,\,\,\,\,\text{,}\label{2.20}%
\end{equation}
\begin{equation}
\frac{\omega_{0}}{l_{u0}}=\frac{2\cdot\pi}{\lambda_{0}}%
\,\,\,\,\,\,\,\,\,\text{,\thinspace\thinspace\thinspace\thinspace
\thinspace\thinspace\thinspace\thinspace\thinspace\thinspace\thinspace
\thinspace\thinspace\thinspace\thinspace\thinspace\thinspace}\lambda
_{0}=\text{ const.,}\label{2.21}%
\end{equation}
we have
\begin{equation}
\lambda=\lambda_{0}\cdot exp\{-\int[H\mp\frac12\cdot(\nabla_{u}g)(\widetilde
{n}_{\perp},\widetilde{n}_{\perp})]\cdot d\tau\}\,\,\,\,\,\text{.}\label{2.22}%
\end{equation}

Therefore, the length $\lambda$ of the standard periodic signal will decrease
in the proper frame of the standard emitter (observer) if
\begin{equation}
\overline{V}(\widetilde{n}_{\perp},\widetilde{n}_{\perp})=H\mp\frac
12\cdot(\nabla_{u}g)(\widetilde{n}_{\perp},\widetilde{n}_{\perp}%
)>0\,\,\,\text{,\thinspace\thinspace\thinspace\thinspace\thinspace
\thinspace\thinspace\thinspace\thinspace\thinspace}\lambda<\lambda_{0}\text{
\thinspace\thinspace\thinspace\thinspace,}\label{2.23}%
\end{equation}
and $\lambda$ will increase when
\begin{equation}
\overline{V}(\widetilde{n}_{\perp},\widetilde{n}_{\perp})=H\mp\frac
12\cdot(\nabla_{u}g)(\widetilde{n}_{\perp},\widetilde{n}_{\perp}%
)<0\,\,\,\,\,\,\,\,\text{,\thinspace\thinspace\thinspace\thinspace
\thinspace\thinspace\thinspace\thinspace\thinspace}\lambda>\lambda
_{0}\,\,\,\,\,\,\text{.}\label{2.24}%
\end{equation}

If
\begin{equation}
\overline{V}(\widetilde{n}_{\perp},\widetilde{n}_{\perp})=H\mp\frac
12\cdot(\nabla_{u}g)(\widetilde{n}_{\perp},\widetilde{n}_{\perp}%
)=0\,\,\,\,\,\,\,\,\text{,}\label{2.25}%
\end{equation}
then
\[
\lambda=\lambda_{0}=\text{ const.}
\]

\subsection{Relation between the variation of the frequency and the variation
of the absolute value of a standard periodic signal}

From the equation for $l_{u}$%
\begin{equation}
\frac12\cdot\frac{dl_{u}^{2}}{d\tau}=\pm[g(u,a)+\frac12\cdot(\nabla
_{u}g)(u,u)]\label{2.26}%
\end{equation}
we can find the expression for $l_{u}$%
\begin{align*}
\frac{dl_{u}^{2}}{d\tau}  & =\pm2\cdot[g(u,a)+\frac12\cdot(\nabla
_{u}g)(u,u)]=\pm2\cdot\widetilde{V}\text{ \thinspace\thinspace\thinspace
\thinspace\thinspace\thinspace\thinspace\thinspace,}\\
\widetilde{V}  & =g(u,a)+\frac12\cdot(\nabla_{u}g)(u,u)\text{ \thinspace
\thinspace\thinspace\thinspace\thinspace\thinspace.}%
\end{align*}

Then
\begin{equation}
l_{u}=(l_{u0}^{2}\pm2\cdot\int\widetilde{V}\cdot d\tau)^{1/2}%
\,\,\,\,\,\,\,\,\,\,\text{,\thinspace\thinspace\thinspace\thinspace
\thinspace\thinspace\thinspace\thinspace\thinspace\thinspace\thinspace
\thinspace\thinspace\thinspace}l_{u0}^{2}=\,\text{const.\thinspace}>0\text{
\thinspace\thinspace\thinspace\thinspace.}\label{2.27}%
\end{equation}

If we now substitute the expression for $l_{u}$ in the expression for $\omega$
we can find the general relation for the variation of the frequency $\omega$
under the variation of the absolute value $l_{u}$ of a standard periodic
signal with the proper time of the emitter
\begin{align}
\omega & =\omega_{0}\cdot(1\pm\frac2{l_{u0}^{2}}\cdot\int[g(u,a)+\frac
12\cdot(\nabla_{u}g)(u,u)]\cdot d\tau)^{1/2}\cdot\nonumber\\
& \cdot exp(\int[H\mp\frac12\cdot(\nabla_{u}g)(\widetilde{n}_{\perp
},\widetilde{n}_{\perp})]\cdot d\tau)\,\,\,\,\,\,\text{.}\label{2.28}%
\end{align}

As a final results, for the variations of the absolute value $l_{u}$, the
frequency $\omega$, and the length $\lambda$ of a standard periodic signal
with the proper time of a standard emitter we obtain the relations
\begin{align}
l_{u}  & =(l_{u0}^{2}\pm2\cdot\int[g(u,a)+\frac12\cdot(\nabla_{u}g)(u,u)]\cdot
d\tau)^{1/2}\,\,\,\,\,\,\,\,\,\,\text{,\thinspace\thinspace\thinspace
\thinspace\thinspace\thinspace\thinspace\thinspace\thinspace}\tag{A}\\
\text{\thinspace\thinspace\thinspace\thinspace\thinspace}l_{u0}^{2}  &
=\,\text{const.\thinspace}>0\text{ \thinspace,}\nonumber
\end{align}
\begin{align}
\omega & =\omega_{0}\cdot(1\pm\frac2{l_{u0}^{2}}\cdot\int[g(u,a)+\frac
12\cdot(\nabla_{u}g)(u,u)]\cdot d\tau)^{1/2}\cdot\nonumber\\
& \cdot exp(\int[H\mp\frac12\cdot(\nabla_{u}g)(\widetilde{n}_{\perp
},\widetilde{n}_{\perp})]\cdot d\tau)\,\,\,\,\,\,\,\,\text{,}\tag{B}%
\end{align}
\begin{align}
\lambda & =\lambda_{0}\cdot exp\{-\int[H\mp\frac12\cdot(\nabla_{u}%
g)(\widetilde{n}_{\perp},\widetilde{n}_{\perp})]\cdot d\tau
\}\,\,\,\,\,\,\text{,}\tag{C}\\
\lambda_{0}  & =\text{ const.}\nonumber
\end{align}

\textit{Special case:} $\overline{U}_{n}$-, $\overline{V}_{n}$-, $U_{n}$-, and
$V_{n} $-spaces. For these types of spaces $\nabla_{u}g=0$, $\overline{V}=\mp
d$, $\overline{V}(\widetilde{n}_{\perp},\widetilde{n}_{\perp})=H$,
$\widetilde{V}=g(u,a)$.
\[
l_{u}=(l_{u0}^{2}\pm2\cdot\int g(u,a)\cdot d\tau)^{1/2}%
\,\,\,\,\text{,\thinspace\thinspace\thinspace\thinspace\thinspace
\thinspace\thinspace\thinspace}l_{u0}^{2}=\,\text{const.\thinspace}>0\text{
\thinspace,\thinspace\thinspace\thinspace}
\]
\[
\omega=\omega_{0}\cdot(1\pm\frac2{l_{u0}^{2}}\cdot\int g(u,a)\cdot
d\tau)^{1/2}\cdot exp(\int H\cdot d\tau)\,\,\,\,\,\,\,\,\text{,}
\]
\[
\lambda=\lambda_{0}\cdot exp(-\int H\cdot d\tau
)\,\,\,\,\,\,\,\,\,\,\,\text{,\thinspace\thinspace\thinspace\thinspace
\thinspace\thinspace\thinspace\thinspace\thinspace\thinspace\thinspace
\thinspace\thinspace\thinspace\thinspace}\lambda_{0}=\text{ const.}\,
\]

\textit{Special case:} $\overline{U}_{n}$-, $\overline{V}_{n}$-, $U_{n}$-, and
$V_{n} $-spaces. $\nabla_{u}g=0$, $\overline{V}=\mp d$, $\overline
{V}(\widetilde{n}_{\perp},\widetilde{n}_{\perp})=H=H_{0}=$ const.,
$\widetilde{V}=g(u,a)$.
\[
\mp d(\widetilde{n}_{\perp},\widetilde{n}_{\perp})=H_{0}=\text{
const.\thinspace\thinspace\thinspace\thinspace,}
\]
\[
\mp[\frac1{n-1}\cdot\theta\mp\sigma(\widetilde{n}_{\perp},\widetilde{n}%
_{\perp})]\,=H_{0}\,\,\,\,\,\,\text{,}
\]
\[
\sigma(\widetilde{n}_{\perp},\widetilde{n}_{\perp})=\pm\frac1{n-1}\cdot
\theta+H_{0\text{ \thinspace\thinspace\thinspace\thinspace}}\,\,\,\text{,}
\]
\[
l_{u}=(l_{u0}^{2}\pm2\cdot\int g(u,a)\cdot d\tau)^{1/2}%
\,\,\,\,\text{,\thinspace\thinspace\thinspace\thinspace\thinspace
\thinspace\thinspace\thinspace}l_{u0}^{2}=\,\text{const.\thinspace}>0\text{
\thinspace,\thinspace}
\]
\begin{align*}
\omega & =\omega_{0}\cdot(1\pm\frac2{l_{u0}^{2}}\cdot\int g(u,a)\cdot
d\tau)^{1/2}\cdot exp(H_{0}\cdot\tau)\,\,\,\,\,\,\,\,\,\text{,}\\
\omega_{0}  & =\text{ const.,}\,
\end{align*}
\[
\lambda=\lambda_{0}\cdot exp(-H_{0}\cdot\tau)\,\,\,\,\,\,\,\,\text{,\thinspace
\thinspace\thinspace\thinspace\thinspace\thinspace\thinspace\thinspace
\thinspace\thinspace\thinspace\thinspace}\lambda_{0}=\text{ const.}
\]

\textit{Special case: }$\overline{U}_{n}$-, $\overline{V}_{n}$-, $U_{n}$-, and
$V_{n} $-spaces. $\nabla_{u}g=0$, $\overline{V}=\mp d$, $\overline
{V}(\widetilde{n}_{\perp},\widetilde{n}_{\perp})=H=H_{0}=$ const.,
$\widetilde{V}=g(u,a)=0$.
\[
l_{u}=l_{u0}\,\,\,\text{,\thinspace\thinspace\thinspace\thinspace
\thinspace\thinspace\thinspace\thinspace}l_{u0}=\,\text{const.\thinspace
}>0\,\,\,\,\,\,\,\text{,}
\]
\[
\omega=\omega_{0}\cdot exp(H_{0}\cdot\tau)\,\,\,\,\,\,\,\,\,\text{,\thinspace
\thinspace\thinspace\thinspace\thinspace\thinspace\thinspace\thinspace
\thinspace}H_{0}=\text{const.,\thinspace}
\]
\[
\lambda=\lambda_{0}\cdot exp(-H_{0}\cdot\tau)\,\,\,\,\,\,\,\,\text{,\thinspace
\thinspace\thinspace\thinspace\thinspace\thinspace\thinspace\thinspace
\thinspace\thinspace\thinspace\thinspace}\lambda_{0}=\text{ const.}
\]

Therefore, if $g(u,a)=0$ and $H=H_{0}$ the absolute value $l_{u0}$ of the
velocity of a periodic signal of a standard emitter is a constant quantity
along the world line of the emitter. The frequency $\omega$ as well as the
length $\lambda$ of the signal are depending exponentially on the time
parameter $\tau$. In general relativity in $V_{n}$-spaces ($n=4$), a standard
periodic light signal of a standard emitter will have a constant value
$l_{u0}$ of its velocity if the acceleration $a$ is orthogonal to the vector
field $u$ or if the world line of the standard emitter is a geodesic
trajectory $(a=0)$ in space-time.

\textit{Special case:} Shear free and expansion free $(\overline{L}_{n}%
,g)$-spaces. $\sigma=0$, $\theta=0$. For shear free and expansion free
$(\overline{L}_{n},g)$-spaces $H=0$ and we have the relations
\begin{align*}
l_{u}  & =(l_{u0}^{2}\pm2\cdot\int[g(u,a)+\frac12\cdot(\nabla_{u}g)(u,u)]\cdot
d\tau)^{1/2}\,\,\,\,\,\,\,\,\,\,\text{,\thinspace\thinspace}\\
\text{\thinspace}l_{u0}^{2}  & =\,\text{const.\thinspace}>0\text{ \thinspace,}%
\end{align*}
\begin{align*}
\omega & =\omega_{0}\cdot(1\pm\frac2{l_{u0}^{2}}\cdot\int[g(u,a)+\frac
12\cdot(\nabla_{u}g)(u,u)]\cdot d\tau)^{1/2}\cdot\\
& \cdot exp(\mp\frac12\cdot\int(\nabla_{u}g)(\widetilde{n}_{\perp}%
,\widetilde{n}_{\perp})\cdot d\tau)\,\,\,\,\,\,\,\,\text{,}%
\end{align*}
\[
\lambda=\lambda_{0}\cdot exp(\pm\frac12\cdot\int(\nabla_{u}g)(\widetilde
{n}_{\perp},\widetilde{n}_{\perp})d\tau)\text{ \thinspace\thinspace
\thinspace\thinspace\thinspace.}\,
\]

\section{Variation of the velocity and the frequency of periodic signals of a
standard oscillator (clock)}

Let us now consider the change of the frequency of periodic signals of a
standard oscillator used as a standard clock in the frame of reference of an
observer. The period of the oscillator, $T=2\cdot\pi/\omega$, is calibrated in
some standard oscillator. If the frequency $\omega$ is changing along the
world line of the oscillator then the period $T$ is also changing in the
corresponding form
\begin{align}
T  & =\frac{2\cdot\pi}\omega=T_{0}\cdot(1\pm\frac2{l_{u0}^{2}}\cdot
\int[g(u,a)+\frac12\cdot(\nabla_{u}g)(u,u)]\cdot d\tau)^{-1/2}\cdot\nonumber\\
& \cdot exp(-\int[H\mp\frac12\cdot(\nabla_{u}g)(\widetilde{n}_{\perp
},\widetilde{n}_{\perp})]\cdot d\tau)\,\,\,\,\,\,\,\,\text{,}\label{3.1}\\
T_{0}  & =\frac{2\cdot\pi}{\omega_{0}}=\text{ const.}\nonumber
\end{align}

Therefore, a standard oscillator (clock) would not change its frequency and
period if and only if the following conditions are fulfilled
\begin{align}
g(u,a)+\frac12\cdot(\nabla_{u}g)(u,u)  & =0\,\,\,\,\,\,\,\text{,}\label{3.2}\\
H\mp\frac12\cdot(\nabla_{u}g)(\widetilde{n}_{\perp},\widetilde{n}_{\perp})  &
=0\,\,\,\,\,\,\text{.}\label{3.3}%
\end{align}

These conditions are in generally not fulfilled even in the Einstein theory of
gravitation, where the above relations should also be valid in their special
forms
\begin{align}
g(u,a)  & =0\,\,\,\,\,\,\,\text{,}\label{3.4}\\
H  & =0\,\,\,\,\,\,\text{.}\label{3.5}%
\end{align}

In all cosmological models in general relativity with Hubble function
$H=H(\tau)$ different from zero and $g(u,a)=0$ a standard clock will move in
space-time with a period $T$ obeying the condition
\begin{equation}
T=\frac{2\cdot\pi}\omega=T_{0}\cdot exp(-\int H\cdot d\tau
)\,\,\,\,\,\,\,\,\,\,\text{.}\label{3.6}%
\end{equation}

Furthermore, the condition $g(u,a)=0$ is considered as a corollary of the
assumption of the constant value $l_{u}=c=$ const. or $l_{u}=1$ of the speed
of light. There is no unique physical argument for the last assumption if a
theory of gravitation has to describe the behavior of physical systems
including the speed of propagation of their interactions on the basis of its
own structures.

\section{Conclusions}

In the present paper the variation of the absolute value and the frequency of
a periodic signal sent by a standard emitter is considered. The obtained
results contradict with the general belief that a standard emitter does not
change the frequency of its periodic signals on its world line considered also
as the world line of an observer (detector) moving together with the standard
emitter. The periodic signals of a standard emitter could have constant period
and frequency only under very specific conditions. They should be fulfilled if
an exact comparison with incoming signals is required. In all other cases
there is a difficult task for an observer to make conclusions about the real
variations of the frequency of an incoming periodic signal in his proper frame
of reference without knowing the exact kinematic characteristics of the
tangent vector field of his world line. If the kinematic characteristics
related to the notion of relative velocity of an observer (shear, rotation,
and expansion velocities) are given (or found by the use of a theoretical
scheme) then the variation of the frequency of a periodic signal, incoming to
the observer, could be easily determined. The possible solutions of this
problem are worth to be investigated in an other paper. The same conclusions
are valid if we try to compare the periods of different clocks on the basis of
a standard clock (oscillator, emitter) moving with an observer.

\end{document}